\documentstyle[12pt,psfig,titlepage]{article}
\def\aj{\ref@jnl{AJ}}                   
\def\araa{\ref@jnl{ARA\&A}}             
\def\apj{\ref@jnl{ApJ}}                 
\def\apjl{\ref@jnl{ApJ}}                
\def\apjs{\ref@jnl{ApJS}}               
\def\applopt{\ref@jnl{Appl.Optics}}     
\def\apss{\ref@jnl{Ap\&SS}}             
\def\aap{\ref@jnl{A\&A}}                
\def\aapr{\ref@jnl{A\&A~Rev.}}          
\def\aaps{\ref@jnl{A\&AS}}              
\def\azh{\ref@jnl{AZh}}                 
\def\baas{\ref@jnl{BAAS}}               
\def\jrasc{\ref@jnl{JRASC}}             
\def\memras{\ref@jnl{MmRAS}}            
\def\mnras{\ref@jnl{MNRAS}}             
\def\pra{\ref@jnl{Phys.Rev.A}}          
\def\prb{\ref@jnl{Phys.Rev.B}}          
\def\prc{\ref@jnl{Phys.Rev.C}}          
\def\prd{\ref@jnl{Phys.Rev.D}}          
\def\prl{\ref@jnl{Phys.Rev.Lett}}       
\def\pasp{\ref@jnl{PASP}}               
\def\pasj{\ref@jnl{PASJ}}               
\def\qjras{\ref@jnl{QJRAS}}             
\def\skytel{\ref@jnl{S\&T}}             
\def\solphys{\ref@jnl{Solar~Phys.}}     
\def\sovast{\ref@jnl{Soviet~Ast.}}      
\def\ssr{\ref@jnl{Space~Sci.Rev.}}      
\def\zap{\ref@jnl{ZAp}}

\def\deg{\hbox{$^\circ$}}

\def\la{\mathrel{\hbox{\rlap{\hbox{\lower4pt\hbox{$\sim$}}}\hbox{$<$}}}}
\def\ga{\mathrel{\hbox{\rlap{\hbox{\lower4pt\hbox{$\sim$}}}\hbox{$>$}}}}

\newcommand{\lapprox }{{\lower0.8ex\hbox{$\buildrel <\over\sim$}}}
\newcommand{\gapprox }{{\lower0.8ex\hbox{$\buildrel >\over\sim$}}}
 
\newcommand{\hii}{\mbox {H\,{\sc ii}}}                  
\newcommand{\hi}{\mbox {H\,{\sc i}}}                  

\def\h2o{H$_2$O}

\def\mone{$^{-1}$}
\def\mtwo{$^{-2}$}
\def\mthree{$^{-3}$}

\def\h2o{H$_2$O}
\def\d21{D_{21}}

\def\n11{n_{11}}
\def\n10{n_{10}}
\def\t400{T_{400}}
\def\f1{f_{(-2)}}

\def\e4{\epsilon_{(-4)}}

\title{A Direct Image of the Obscuring Disk Surrounding the Active Galactic
  Nucleus of NGC~1068}
\author{
Jack F. Gallimore\\
{\em Max-Planck-Institut f\"ur extraterrestrische Physik,}\\ 
{\em Postfach 1603, D-85740 Garching b. M\"unchen, Germany}, \\
\\
Stefi A. Baum and Christopher P. O'Dea \\
{\em Space Telescope Science Institute,}\\
{\em  3700 San Martin Dr., Baltimore, MD 21218, USA}
}

\date{\today}

\begin{document}

\maketitle

\baselineskip=24pt

{\bf Active galactic nuclei (AGNs) are commonly thought to be powered
  by the exchange of gravitational energy for thermal energy in a
  compact accretion disk surrounding a massive black hole
  \cite{Rees84,Gunn79}. Such disks are also 
  necessary to collimate powerful radio jets \cite{BBR84}.  The
  unifying schemes for AGN classification further propose that gas
  fuelling the AGN may obstruct our sight-line, hiding some AGNs from
  direct view; the popular model is a parsec-scale
  (1~pc $\bf= 10^{18.5}$~cm) disk of dense molecular gas
  \cite{Lawrence87}. Evidence for such disks has been mostly indirect,
  since the angular size is much smaller than the resolution of
  conventional telescopes. We report the
  first direct images of a pc-scale disk of ionised gas located within the
  nucleus of NGC~1068, the archetype of obscured AGNs.  The disk is
  viewed nearly edge-on, and individual clouds within the ionised disk
  are opaque to high-energy radiation, consistent with the unifying
  schemes model. In projection, the disk and AGN axes align, and so the
  ionised gas disk may be considered to trace the outer regions of the
  long-sought accretion disk.}

The nucleus of the galaxy NGC~1068 hosts the archetypal example of an
obscured AGN. The popular model for the obscuring medium is a
parsec-scale, molecular disk surrounding the AGN \cite{AM85}, perhaps
ultimately feeding an accretion disk \cite{KB88,PV95}. One difficulty
for observational tests has been that the location of the obscured,
central ionising source is unknown.  It has been argued on several
grounds that the radio source S1 marks the location of the hidden
AGN in NGC~1068 \cite{GBO96,Muxlow96}. Located at the
southern end of the arcsecond-scale radio jet, S1 is an unusual radio
source in two respects. Firstly, in contrast with the rest of the
radio jet, its radio spectrum is relatively flat (spectral
index $\alpha = +0.3$; $S_{\nu}\propto \nu^{\alpha}$ \cite{GBO96}),
and secondly, it is a source of \h2o\ and OH maser emission
\cite{GBOBP96}, which distinguishes regions of peculiarly warm
(1000~K) and dense ($\ga 10^8$ molecules cm\mthree) molecular gas. We
argued that S1 might trace emission from molecular clouds defining the
inner surface of the proposed obscuring disk, whose surfaces would be
exposed directly to the central X-ray source and are therefore hot and
highly ionised \cite{GBO96}.

There are two main predictions for the high resolution observations
presented here. The obvious prediction is that S1 should resolve into
a pc-scale, linear radio structure, tracing the profile of an edge-on
disk or `torus' projected onto the sky, and located within the warm,
molecular disk mapped in part by \h2o\ masers \cite{GBOBP96}.
Secondly, the mean surface brightness of S1, in temperature units
corresponding to an equivalent blackbody radiator (brightness
temperature), should be $T_b \sim 10^6$~K for scattering-diffused
emission or thermal free-free emission \cite{GBO96}.

To test these predictions, we have imaged the subarcsecond jet of
NGC~1068 using the 10-station Very Large Baseline Array (VLBA),
augmented by the phased, 26-element Very Large Array (VLA). The new
images are displayed in Fig.~\ref{images}. A single, deep (8.8 hrs
on-source) integration was obtained at 8.4 GHz only.  The observations
and data reduction followed standard techniques with exceptions as
follows. On continental-scale baselines, NGC~1068 is not sufficiently
bright at 8~GHz to calibrate interferometric fringes within averaging
times comparable to the atmospheric coherence time. Instead, we
measured fringe-rate and -delay corrections from short scans of the
nearby calibrator source 0237$-$027. Less than 0.2\% of the data (one
out of a total of nearly 600 baseline-hours) were affected by phase
rotations resulting from fringe solution ambiguities, and the radio
sources S1, C, and NE were clearly detected on the initial maps. The
resulting, fringe-calibrated data were coherent over sufficiently long
intervals to permit self-calibration, and so we removed the residual
phase wraps using five iterations of phase-only self-calibration. In
order to focus specifically on the nuclear emission, the VLBA images
of NE and C will be presented elsewhere. We focus instead on the `Hot
Zone' (HZ), the brighter, central region of S1, and
also defer discussion of fainter radio emission to future work.

The HZ comprises nine distinguishable compact sources, each of total
flux density $S_{\nu} \la 0.65$~mJy (1~mJy $=10^{-26}$ ergs s\mone\ 
cm\mtwo\ Hz\mone), embedded in diffuse emission. These observations
only marginally resolve the individual compact sources. Based on
Gaussian model fits and image moment analysis, the deconvolved source
sizes are typically $\sim 1$~milliarcsecond (mas), or $\sim 0.07$~pc at the
distance of NGC~1068. These measurements are however uncertain owing
to confusion between neighboring sources, blurring due to residual
phase errors, and possible enhancement by deconvolution. 
We also estimated limits on the source sizes
based on inspection of the interference fringes.  Less than half
($\sim 3$~mJy) of the recovered flux of the HZ is detected on
baselines corresponding to angular sizes $< 2 \times 1$~mas; at least
half of the flux from the HZ must therefore arise from structures $\ga
1$~mas in size, consistent with measurements of the synthesized image.
This limit is conservative, since 1--2~mJy worth of mas-scale
components are also detected towards components NE and C.  Any one or
two of these compact sources may be smaller than 1~mas, but this
caveat will not affect the main conclusions.

The compact sources trace a slightly curving line along a position
angle of 110\deg, measured east of north. The HZ is therefore
nearly at right angles with the collimation axes defined by the local
radio jet and polarisation axes, the latter of which describes the
collimation axis for escaping ionising radiation \cite{AHM94}. This
geometry is fully consistent with our prediction that the radio
emission from S1 traces emission not from a streaming jet but rather
from gas in an ionised disk surrounding the AGN. We next consider the
implications of this disk model, making the simplifying assumption
that the HZ gas lies at a common radius, $\sim 0.3$--0.5~pc, from the
AGN.

Opaque synchrotron emission is the conventional explanation for
flat-spectrum radio sources, but the brightness temperatures over the
HZ are too low for synchrotron self-absorption \cite{syncreview}.
There are two likely alternatives, illustrated in Fig.~\ref{schem},
either being a variation on emission from ionised gas in a disk
\cite{GBO96}. The first is synchrotron emission from a
synchrotron-opaque, compact radio source, presumably the AGN, which is
not viewed directly but in reflection by electron-scattering from the
ionised gas disk. 

The limits for this model are set by requiring that the electron
scattering opacity ($\tau_{e}$) must exceed the opacity to free-free
absorption ($\tau_{ff}$), and that the hidden radio source must not be
so luminous that it would have been detected in reflection on larger
scales. The limits for this model are set by requiring that the
electron scattering opacity ($\tau_{e}$) must exceed the opacity to
free-free absorption ($\tau_{ff}$), and that the hidden radio source
must not be so luminous that it would have been detected in reflection
on larger scales. Based on the sensitivities of our radio continuum
images \cite{GBO96} and the reflecting properties of the
electron-scattering mirror \cite{Capetti96}, we estimate that the
hidden radio source can be no brighter than $S_{\bullet} \la 3.5$~Jy.
We estimated limits on the plasma properties by exploring a grid of
$n_e$, $T_e$, and cloud thicknesses $l$ (and the corresponding value
$S_{\bullet}$ appropriate for a given $\tau_e$) and rejecting those
values where $\tau_{ff} > 0.5\tau_{e}$ and $S_{\bullet} > 3.5$~Jy.  We
find that the reflection model can be satisfied for electron
temperatures $T_e \ga 10^{6.7}$~K, electron densities $10^{6.2} \la
n_e < 10^{6.6}$~electrons cm\mthree, and ionised cloud thicknesses
$0.007 \la l \la 0.07$~pc (the upper limit set by the measured sizes).
We also estimate that, assuming that thermal absorption is negligible,
the flux density of any hidden compact radio source must be $0.8 \la
S_{\bullet} \la 3.5$~Jy.

The second model is direct, thermal free-free emission from ionised
gas inside the obscuring disk. Appropriate for the integrated radio
spectrum \cite{Muxlow96,GBOP96}, we assume for this thermal model a mean
opacity of $\tau_{ff}(8.4{\rm\ GHz}) = 0.5$ through the Hot Zone
plasma. Using the free-free opacity approximations of Mezger and
Henderson \cite{MH67}, we estimate $10^{6.5} \la T_e \la 
10^{6.8}$~K and $n_e \ga 10^{6.8}\ {\rm cm^{-3}}\ (T_e / 10^7\ {\rm
  K})^{1.35} (l/0.07\ {\rm pc})^{0.5}$. 

The plasma conditions in either the thermal or reflection models are
plausible given the extreme environment. For comparison, particle
densities in the molecular region of the disk are estimated to be
$n_{H_2} \approx 10^8$~molecules cm\mthree\ \cite{PV95,NMC94}, and
photoionisation heating can drive $T_e$ up to the limit bound by
inverse Compton cooling, $T_C \sim 10^7$--$10^8$~K \cite{PV95,KMT81}.
On the other hand, it is not clear how such a dense medium can remain
heated to temperatures so near the Compton limit. For instance,
photoionisation heating of dense plasmas can support only $T_{cool}
\sim 10^4$--$10^6$~K \cite{PV95,KB86}, unless the ionising spectrum is
much harder (more luminous in X-rays) and overall more luminous than
current estimates \cite{Pier94}.  A promising alternative is heating
by mixing with a hot, intercloud medium \cite{RF95}, a $T_{hot} \sim
10^7$--$10^8$~K plasma proposed to confine optical emission line
clouds in the nuclear environment \cite{KMT81}.  The temperatures in
the mixed plasma would be $T_e \approx \sqrt{T_{cool} T_{hot}}$
\cite{BF91}, or $T_e \sim 10^{6.5}$~K, to within factors of a few.  In
addition, fast shocks, such as those driven by winds or the radio jet,
or internal shocks arising at cloud-cloud collisions, might at least
transiently support such high temperatures. Understanding the energy
budget of the HZ will be a challenge for follow-up research.


Relevant specifically to AGN unifying schemes, the column density
through the HZ clouds may be as high as $n_e l \sim 10^{24}$~cm\mtwo,
sufficient to absorb virtually all of the incident X-ray emission from
the AGN \cite{MMW92}. This result and that the disk is viewed nearly
edge-on argue that the HZ traces ionised gas lying within the
obscuring disk of NGC~1068. One difficulty is that the covering
fraction of the compact sources, $\sim 5$--10\%, is much smaller than
required generally to explain the fraction of directly viewed AGNs
\cite{Lawrence91}. However, the model can be reconciled if the
geometric thickness of the obscuring medium increases with radius
\cite{Efstathiou95}, and the HZ traces only emission from disc
material nearest the AGN.  Alternatively, the obscuring disk might
instead be thin but highly warped \cite{Sanders89}, and again the HZ
marks only the centermost region of the disk.

Turning to the broad-band properties of the disk, the HZ plasma must
also be a source of line and continuum emission up to soft X-ray
energies. To determine whether the optical--X-ray spectrum of the HZ
might be distinguished from neighboring emission line regions and the
AGN proper, we modelled the HZ spectrum using the {\sc cloudy}
photoionisation code \cite{Ferland93}, programmed to emulate a cooling
plasma with the properties of the HZ, and normalised to the observed
radio flux.  We find that, in broad agreement with Pier and Voit
\cite{PV95}, the HZ contributes $\la 10\%$ to the observed
optical--UV emission lines of NGC~1068 and $\la 1\%$ to the
optical--UV continuum. On the other hand, the HZ should contribute
significantly to the soft X-ray spectrum of the nucleus were the AGN
viewed along an unobscured sight-line. We estimate that free-free
emission from the HZ may contribute $\ga 10\%$ of the AGN
continuum in soft X-rays (photon energies $\sim 1$--2 keV). Moreover, the
predicted soft X-ray line emission exceeds that observed toward
NGC~1068 \cite{Ueno94} by a factor of $\ga 100$. This observed
dimunition is approximately that expected for {\em obscured} X-ray
emission viewed only in reflection \cite{Pier94}. Therefore, this
result can be reconciled if the HZ is also heavily obscured over
optical--soft X-ray wavebands. The implication is that, if the AGN
unifying schemes hold generally, the disks surrounding unobscured AGNs
should be luminous soft X-ray emission line sources. To our knowledge,
there has as yet been no analysis of the soft X-ray line emission from
unobscured AGNs (i.e., Seyfert 1 AGNs). The detection of soft X-ray
lines characteristic of a $10^6$--$10^7$~K plasma in unobscured AGNs
would lend self-consistency to the obscuring disk model.

The present observation is the first direct image of a pc-scale,
ionised gas disk surrounding an AGN. Simple models for the radio
emission further provide the first direct estimates of the physical
properties of a pc-scale disk, and the results are consistent with the
predictions of AGN unifying schemes. However, we also find that
photoionisation heating is insufficient to support the high plasma
temperatures in the disk, and so the challenge remains to model the
energy budget in accord with the observed radio emission. Equally
important is how the HZ might fit into the standard, infall
model for AGNs \cite{Gunn79}.  The HZ is oriented nearly at right
angles to the radio jet. The observed orientation suggests that,
within the HZ, internal, viscous dissipation drives the fueling of the
AGN rather than external torques. In this regard, the HZ may be
considered to define the outer extent of the long-sought accretion
disk powering the AGN.

\vspace{0.5truein}

\noindent Correspondence and requests for materials to J.F.G.
    (e-mail: jfg@hethp.mpe-garching.mpg.de).

\noindent{\bf Acknowledgements.} The VLBA and VLA are operated by the
    National Radio Astronomy Observatory which is operated by
    Associated Universities, Inc., under cooperative agreement with
    the National Science Foundation. J.F.G. received support from a
    Collaborative Visitor's Grant from the Space Telescope Science
    Institute. We also acknowledge useful suggestions from an
    anonymous referee which helped to clarify the text.

\vspace{0.5truein}

\clearpage 

\begin{figure}[h!]
\caption{\protect\baselineskip=24pt New VLBI images of the radio
  component S1 of NGC~1068. The 
  total recovered flux of the HZ is 6.9~mJy, or roughly 60\% of
  the flux anticipated by a power-law interpolation of the 5~GHz and
  22~GHz measurements of Gallimore et al.  \protect\cite{GBOP96}. In
  contrast, less than a total of $\sim 1$~mJy arises from compact
  structures lying outside the Hot Zone (HZ) but within S1. Owing to an
  instability in the deconvolution algorithm used to produce this
  image, some of the compact sources may be artificially enhanced at
  the expense of the diffuse emission \protect\cite{CLEANref}. The compact
  sources are nevertheless real, since they are also distinguishable
  on the unprocessed images. {\sl Upper plot:} Naturally weighted
  image of S1; the 
  beam size (FWHM), indicated by the darkened ellipse, is $2.5 \times
  1.4$~mas.  We have marked and labeled the HZ and the local jet
  axis towards radio jet component C. Note that, in projection, the
  extent of the HZ and the direction of the radio jet are at
  right angles to each other, 
  suggesting a common symmetry axis. Scaled
  logarithmically, the contour levels are $\pm 0.10\ (2.5\sigma)$, 0.22,
  0.35, 0.47, and 0.59 mJy beam\mone, or $\pm 0.49$, 1.1, 1.7, 2.3,
  and 2.9 in brightness temperature units of $10^6$~K. {\em Lower
    plot:} Uniformly weighted image of the HZ; the beam size
  (FWHM) is $2.3\times 1.1$~mas. The contour levels are $\pm 0.16\ 
  (2.5\sigma)$, 0.24, 0.36, and 0.54 mJy beam\mone, or $\pm 1.1$, 1.6,
  2.5, $3.7 \times 10^6$~K.}
\label{images}
\end{figure}
\clearpage

\centerline{\psfig{file=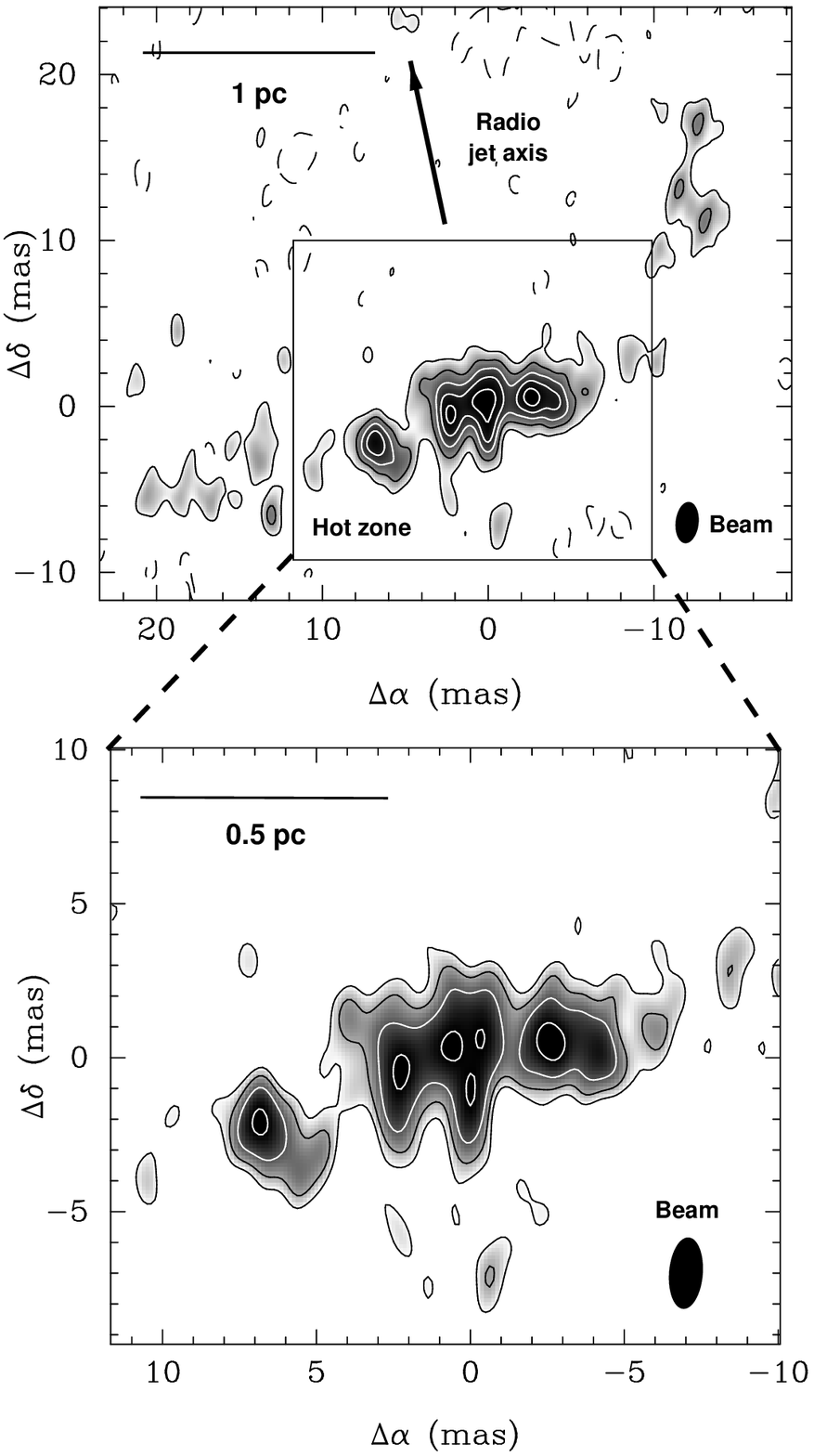,height=6.5truein}}

\clearpage
\begin{figure}[h!]
\caption{\protect\baselineskip=24pt  Model schematics of the
  AGN of NGC~1068 and its environs. {\em Left Panel:} A cartoon
  depicting the obscuring disk as viewed along our sight-line.  The AGN
  is hidden from direct view, but we see reprocessed and scattered
  emission from the surrounding disk  (illuminated material at the
  center of the cartoon). {\em Right Panel:} A plan view of the same,
  but which also 
  illustrates the relative locations of the $10^6$~K Hot Zone,
  detected in these observations, the warm transition zone, traced by
  \h2o\ maser emission and \hi\ absorption \protect\cite{GBOBP96}, and
  an outer, cooler molecular zone, which still eludes direct
  detection.  It has been argued that the innermost region may be
  filled with a hot ($10^8$~K), intercloud medium
  \protect\cite{KMT81}, which might be a source of heat for the
  HZ. This cartoon also illustrates two possible 
  contributions to the observed radio emission: (1) scattered non-thermal
  emission originating at the AGN and (2) direct free-free emission
  from the Hot Zone.}\label{schem} 
\end{figure}
\clearpage

\centerline{\psfig{file=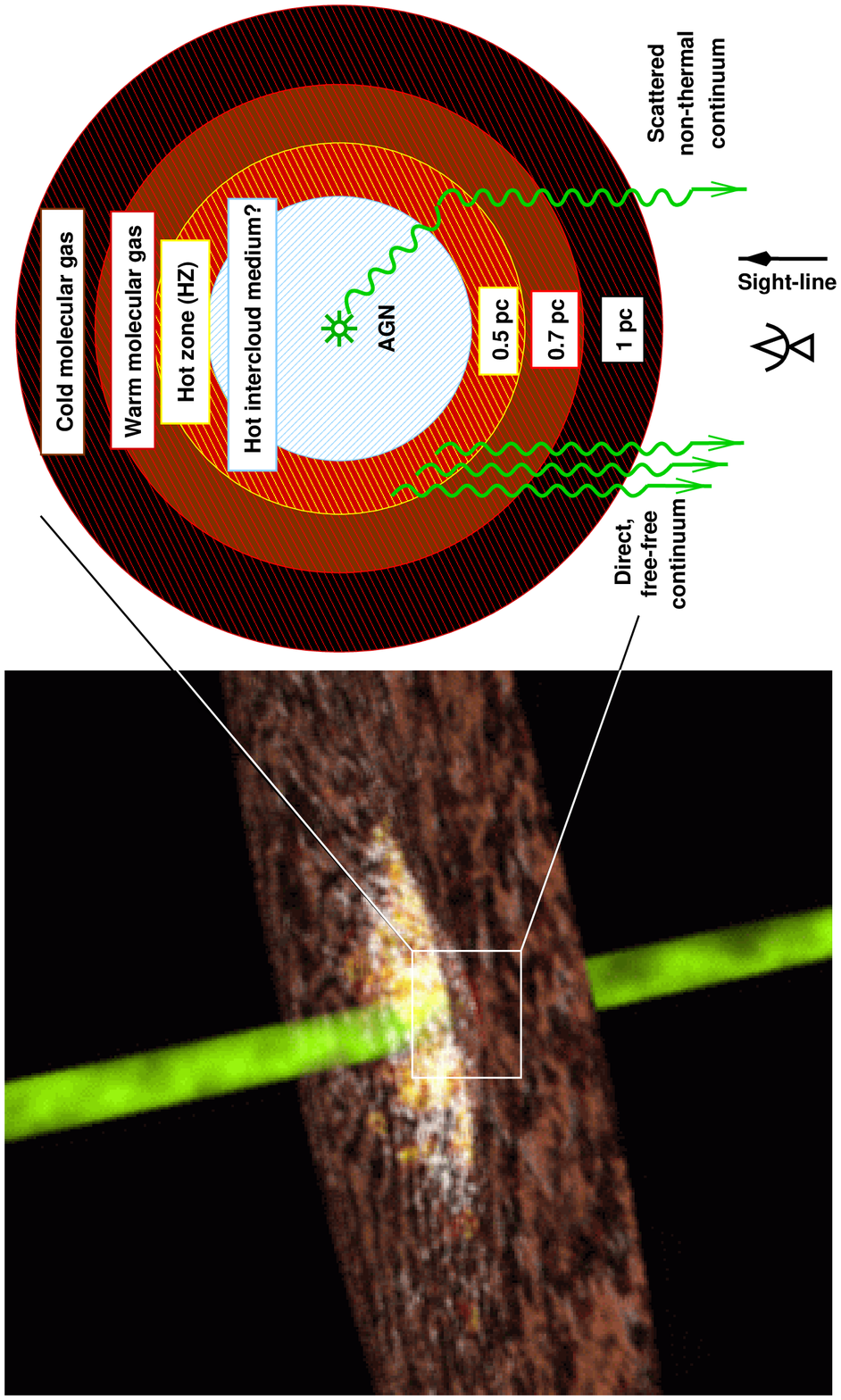,angle=270,width=\textwidth}}


\begin{thebibliography}{99}
%
%
\bibitem{Rees84} Rees, M. J. Black hole models for active galactic
    nuclei. {\em Ann. Reviews Astron. Astrophys.} {\bf 
    22}, 471--506 (1984). 
\bibitem{Gunn79} Gunn, J.E. Feeding the monster -- Gas discs in
    elliptical galaxies. In {\em Active Galactic Nuclei},
  (eds. C. Hazard \& S. Mitton) 213-225  (Cambridge University Press,
    Cambridge, 1979).
\bibitem{BBR84} Begelman, M.C., Blandford, R.D., \& Rees, M. J. Theory
    of extragalactic radio sources. {\em
    Rev. Mod. Phys.} {\bf 56}, 255--351 (1984).
\bibitem{Lawrence87} Lawrence, A. Classification of active galaxies
    and the prospect of a unified phenomenology. {\em
    Publ. Astron. Soc. Pacif.} {\bf 
    99}, 309--334 (1987)
\bibitem{AM85}  Antonucci, R. R. J., \& Miller
    J. S. Spectropolarimetry and the nature of NGC 1068. {\em
    Astrophys. J.} {\bf 
  297}, 621--532 (1985).
\bibitem{KB88}  Krolik, J. H., \& Begelman, M. C. Molecular tori in
    Seyfert galaxies - Feeding the monster and hiding it. {\em
    Astrophys. J.} 
  {\bf 329}, 702--711 (1988). 
\bibitem{PV95} Pier, E.A., \& Voit, G.M. Photoevaporation of dusty
    clouds near active galactic nuclei. {\em Astrophys. J.} {\bf
    450}, 628--637 (1995)  
\bibitem{GBO96}  Gallimore, J. F., Baum, S. A., \& O'Dea, C. P. The
    subarcsecond radio structure in NGC 1068. II. Implications for the
    central engine and unifying schemes. {\em
    Astrophys.  J.} {\bf 464}, 198--211 (1996).
\bibitem{Muxlow96}  Muxlow, T. W. B., Pedlar, A., Holloway, A. J.,
  Gallimore, J. F., \& Antonucci, R. R. J. The compact radio nucleus
    of the Seyfert galaxy NGC 1068. {\em
  Mon. Not. R. Astron. Soc.} {\bf 278}, 854--860 (1996).
\bibitem{GBOBP96} Gallimore, J. F., Baum, S. A., O'Dea, C. P., Brinks,
  E., \& Pedlar, A. \h2o\ and OH masers as probes of the obscuring
    torus in NGC~1068. {\em Astrophys. J.} {\bf 462}, 740--745 (1996)  
\bibitem{GBOP96} Gallimore, J. F., Baum, S. A., O'Dea, C. P., \& 
  Pedlar, A. The subarcsecond radio structure in NGC
    1068. I. Observations and results. {\em Astrophys. J.} {\bf 458},
    136--148 (1996)   
\bibitem{CLEANref} Cornwell, T., \& Braun, R. Deconvolution. In {\em
    Synthesis 
    Imaging in Radio Astronomy} (eds Perley, R.A., Schwab, F.R. \&
    Bridle, A.H.) 167--181 (A.S.P. Conference Series, San Francisco,
    1994). 
\bibitem{AHM94} Antonucci, R.R.J., Hurt, T., \& Miller, J.S. HST
    ultraviolet spectropolarimetry of NGC 1068. {\em
    Astrophys. J.} {\bf 430}, 210--217 (1994)
\bibitem{syncreview} Kellerman, K.I. \& Owen, F. N. Radio galaxies and   
    quasars. In {\em Galactic
    and Extragalactic Radio Astronomy} (eds Verschuur, G.L. \&
    Kellermann, K.I.) 563--600 (Springer, New York, 1989)
\bibitem{Capetti96} Capetti, A., Macchetto, F., Axon, D.J., Sparks,
    W.B., \& Boksenberg, A. Hubble Space Telescope imaging polarimetry
    of the inner nuclear region of NGC 1068. {\em Astrophys. J. Lett.}
    {\bf 452}, L87--L89 (1995).
\bibitem{MH67} Mezger, P.G., Henderson, A.P. Galactic \hii\ regions:
    I. Observations of their continuum radiation at the frequency
    5~GHz. {\em Astrophys. J.} {\bf 
    147}, 471--489 (1967)
\bibitem{NMC94} Neufeld, D.A., Maloney, P.R., \& Conger, S. Water
    maser emission from X-ray-heated circumnuclear gas in active
    galaxies {\em Astrophys. J. Lett.} {\bf 436}, L127--L130 (1994)
\bibitem{KMT81} Krolik, J.H., McKee, C.F. \& Tarter, C.B. Two-phase
    models of quasar emission line regions. {\em 
    Astrophys. J.} {\bf 249}, 422--442 (1981)
\bibitem{KB86} Krolik, J.H., \& Begelman, M.C. An X-ray heated wind in
    NGC 1068. {\em Astrophys. J. Lett.} {\bf
    308}, L55--L58 (1986)
\bibitem{Pier94}  Pier, E. A., Antonucci, R., Hurt, T., Kriss, G., \& Krolik,
  J. The intrinsic nuclear spectrum of NGC 1068. {\em Astrophys. J.}
  {\bf 428}, 124--129 (1994).  
\bibitem{RF95} Reynolds, C.F. \& Fabian, A.C.  Warm absorbers in
  active galactic nuclei. {\em 
    Mon. Not. R. Astron. Soc.} {\bf 273}, 1167--1176 (1995). 
\bibitem{BF91} Begelman, M.C. \& Fabian, A.C. Turbulent mixing layers
  in the interstellar and intracluster medium{\em
    Mon. Not. R. Astron. Soc.} {\bf 244}, 26P--29P.
\bibitem{MMW92}  Mulchaey, J. S., Mushotzky, R. F., \& Weaver,
  K. A. Hard X-ray tests of the unified model for an
  ultraviolet-detected sample of Seyfert 2 galaxies. {\em 
    Astrophys. J.} {\bf 390}, L69--L72 (1992).
\bibitem{Lawrence91} Lawrence, A. The relative frequency of
  broad-lined and narrow-lined active galactic nuclei -- Implications
  for unified schemes. {\em Mon. Not. R. Astron. Soc.} {\bf 252}, 
  586--592 (1991).
\bibitem{Efstathiou95} Efstathiou, A., Hough, J.H., \& Young, S.
    A model for the infrared continuum spectrum of NGC 1068. {\em
  Mon. Not. R. Astron. Soc.} {\bf 277}, 1134--1144 (1995) 
\bibitem{Sanders89} Sanders, D.B., Phinney, E.S., Neugebauer, G.,
    Soifer, B.T., \& Matthews, K. Continuum energy distribution of
    quasars -- Shapes and origins. {\em Astrophys. J.} {\bf 357},
    29--51 (1989)
\bibitem{Ferland93} Ferland, G. {\em HAZY, a Brief Introduction to
    Cloudy}, University of Kentucky Physics and Astronomy Department
    Internal Report (1993)
\bibitem{Ueno94} Ueno, S., Mushotzky, R.F., Koyama, K., Iwasawa, K.,
    Awaki, H., \& Hayashi, I. ASCA observations of NGC 1068. {\em
    Publ. Astron. Soc. J.} {\bf 46} 
    L71--L75 (1994)
\end{thebibliography}
\end{document}